\documentclass[lettersize,journal]{IEEEtran}
\usepackage{amsmath,amsfonts}
\usepackage{algorithmic}
\usepackage{algorithm}
\usepackage{array}
\usepackage[caption=false,font=normalsize,labelfont=sf,textfont=sf]{subfig}
\usepackage{textcomp}
\usepackage{stfloats}
\usepackage{url}
\usepackage{verbatim}
\usepackage{multirow}
\usepackage{graphicx}
\usepackage{booktabs}
\usepackage{url}
\usepackage{cite}
\usepackage[mathletters]{ucs}
\usepackage[utf8x]{inputenc}
\usepackage{ulem}

\usepackage[usenames,dvipsnames,svgnames,table]{xcolor}
\usepackage{array}

\usepackage{xcolor}
\newcommand{\mv}[1]{{\color{green}#1}}
\newcommand{\lm}[1]{{\color{orange}#1}}
\newcommand{\vj}[1]{{\color{olive}#1}}

\newcommand{\trademark}{\textsuperscript{TM}}

\usepackage{pgfplots}
\pgfplotsset{width=\columnwidth,compat=1.9}
\usepackage{pgf-pie} 

\usepgfplotslibrary{external}
\definecolor{color1}{HTML}{c7f9cc}
\definecolor{color2}{HTML}{80ed99}
\definecolor{color3}{HTML}{57cc99}
\definecolor{color4}{HTML}{38a3a5}
\definecolor{color5}{HTML}{22577a}
\definecolor{ceil}{rgb}{0.57, 0.63, 0.81}
\definecolor{cerise}{rgb}{0.87, 0.19, 0.39}
\definecolor{carrotorange}{rgb}{0.93, 0.57, 0.13}

\begin{document}

\title{TinyVers: A Tiny Versatile System-on-chip with State-Retentive eMRAM for ML Inference at the Extreme Edge}

\author{Vikram Jain, Sebastian Giraldo, Jaro De Roose, Linyan Mei, Bert Boons, and Marian Verhelst
\thanks{V. Jain, L. Mei, and M. Verhelst are with the Department of Electrical Engineering - MICAS, KU Leuven, Belgium.}
\thanks{S. Giraldo was with the Department of Electrical Engineering - MICAS, KU Leuven, Belgium. He is now with B12 Consulting, Belgium.}
\thanks{J. De Roose and B. Boons were with the Department of Electrical Engineering - MICAS, KU Leuven, Belgium. They are now with Magics Technologies, Belgium.}
\thanks{$\copyright$ 2023 IEEE.  Personal use of this material is permitted.  Permission from IEEE must be obtained for all other uses, in any current or future media, including reprinting/republishing this material for advertising or promotional purposes, creating new collective works, for resale or redistribution to servers or lists, or reuse of any copyrighted component of this work in other works.}
}




\maketitle

\begin{abstract}
Extreme edge devices or Internet-of-thing nodes require both ultra-low power always-on processing as well as the ability to do on-demand sampling and processing. Moreover, support for IoT applications like voice recognition, machine monitoring, etc., requires the ability to execute a wide range of ML workloads. This brings challenges in hardware design to build flexible processors operating in ultra-low power regime. This paper presents TinyVers, a tiny versatile ultra-low power ML system-on-chip to enable enhanced intelligence at the Extreme Edge. TinyVers exploits dataflow reconfiguration to enable multi-modal support and aggressive on-chip power management for duty-cycling to enable smart sensing applications. The SoC combines a RISC-V host processor, a 17 TOPS/W dataflow reconfigurable ML accelerator, a 1.7 $\mu$W deep sleep wake-up controller, and an eMRAM for boot code and ML parameter retention. The SoC can perform up to 17.6 GOPS while achieving a power consumption range from 1.7 $\mu$W-20 mW. Multiple ML workloads aimed for diverse applications are mapped on the SoC to showcase its flexibility and efficiency. All the models achieve 1-2 TOPS/W of energy efficiency with power consumption below 230 $\mu$W in continuous operation. In a duty-cycling use case for machine monitoring, this power is reduced to below 10 $\mu$W.
\end{abstract}


\section{Introduction}


\IEEEPARstart {E}xtreme edge devices~\cite{ee} or Internet-of-Things (IoT) nodes mostly perform non-vision tasks and can achieve good accuracy, even with small and lightweight neural network (NN) models~\cite{hello_edge}. This is in contrast to more traditional tasks designed for processing image data and contain millions to billions of parameters and operations with high hardware resource demands. Consider the Google voice assistant as an example, which needs only 14 kilo bytes (kB) of NN parameters to run a keyword-spotting application on edge devices~\cite{pete}. The insight 
that not all applications require maximum accuracy, large and complex NN models, has resulted in a new paradigm of ML application development, called tinyML or ML at the extreme edge~\cite{tinyml}. This trend, at its core, has been driven by the requirements imposed by battery-operated, performance- and power-constrained IoT nodes. Most IoT sensor nodes consist of a microcontroller unit (MCU) with a subset of sensors, a memory for storing acquired data, a CPU and a wireless data transceiver. 
The presence of these MCUs for data collection provides opportunities to process data very close to the sensor when the NN model is small, and avoids the high penalty of raw data transmission to more powerful edge or cloud units. 

Yet, this local ML processing, brings several new challenges: 1.) As these nodes are battery-operated, the system is typically severely power or energy constrained requiring ultra-low power operation, with the ability to idle. 2.) the MCU, moreover, has limited compute power and memory space, resulting in a critical trade-off between model size, execution performance and hardware complexity; 
3.) despite the need for efficiency, the system should also be flexible enough to support different classes of NN models across different applications, and 4.) it should have a small footprint. Several hardware for ML have been proposed in the recent literature and can be divided into three main categories: 1) extremely specialized edgeML accelerators designed for ultra-low power operation with little to no flexibility at low performance~\cite{vlsi21,utrail,vocell,laika}, 2) multi-modal edgeML accelerators providing medium level of flexibility with high performance at medium to high power consumption~\cite{isscc22_2, isscc22_1, isscc21, envision, eyeriss}, and, 3) commercial-off-the-shelf (COTS) MCUs delivering higher flexibility but at low performance and medium power consumption~\cite{gapuino,ard,slt}. Most of these hardware designs do not meet all the requirements of an extreme edge device. An exception is Vega~\cite{vega} which presents a complete SoC, however, the specialized accelerator of Vega does not have the flexibility to handle all DNN workloads. Thus, a new class of flexible ultra-low power (ULP) platforms towards extreme edge deployment is needed.

In this context, this work presents TinyVers~\cite{tinyvers}, a highly adaptive SoC platform which significantly enhances the trade-off between energy efficiency and flexibility needed in extreme edge devices, through the use of: A.) a RISC-V processor extended with a flexible ML accelerator (FlexML) with dataflow reconfiguration supporting diverse ML workloads and support for efficient zero-skipping in block structured sparsity and deconvolution; B.) an embedded magnetoresistive random access memory (eMRAM) for non-volatile storage enabling standalone operation with efficient power-down (or idling); C.) a programmable wake-up controller (WuC) supporting different power-on and idle modes to enable both always-on inference as well as on-demand and duty-cycled smart sensing and computation used in typical tinyML IoT applications. The SoC provides users flexibility not only in mapping diverse ML workloads for diverse tinyML applications, but also in supporting various use cases such as duty-cycling and smart sensing. We demonstrate TinyVers' capabilities and improvements over state-of-the-art (SotA) on diverse applications in machine monitoring, anomaly detection, audio signal analysis, and image classification through the use of both deep learning as well as traditional ML workloads.

The rest of the paper is organized as follows. The basics of ML compute kernels is introduced in Section~\ref{sec:background}. Section~\ref{sec:tinyvers} discusses the architecture overview of TinyVers, followed by Section~\ref{sec:accel} providing further details of the FlexML accelerator. Section~\ref{sec:compile} provides details on how the software stack for ML deployment on TinyVers is undertaken. Subsequently, Section~\ref{sec:results} presents the experimental results of mapping different workloads and application use cases. Finally, Section~\ref{sec:sota} compares TinyVers' performance with related works and Section~\ref{sec:con} concludes the paper.  



\section{Algorithmic Background}
\label{sec:background}
ML applications heavily exploit deep neural networks (DNN) with traditional convolutional (CNN) and fully connected (FC) layers. However, a plethora of new NN layer topologies are emerging. Some examples of these are the use of temporal convolutional networks (TCN) used in audio tasks like keyword spotting~\cite{tcn_1,tcn_2,tcn_work}, or auto-encoders (AE) using convolution and deconvolution pairs in machine monitoring and anomaly detection tasks~\cite{cae_1, cae_2,cae}. Morever, also machine learning models not relying on neural network layers are still used in extreme edge IoT nodes, such as support vector machines (SVM)~\cite{svm} used in novelty and anomaly detection applications. The execution efficiency of all these workloads can can be improved with orders of magnitude when deployed on specialized accelerators. Yet, the wide variety in the compute kernels of interest 
complicates their efficient mapping on a single hardware platform. The following subsections deal with the different ML operation characteristics, their categorization into mathematical operations, and their hardware implications.

\subsection{Convolution and Dense Operation}
\begin{figure}[!t]
\centering
\includegraphics[width=0.9\columnwidth]{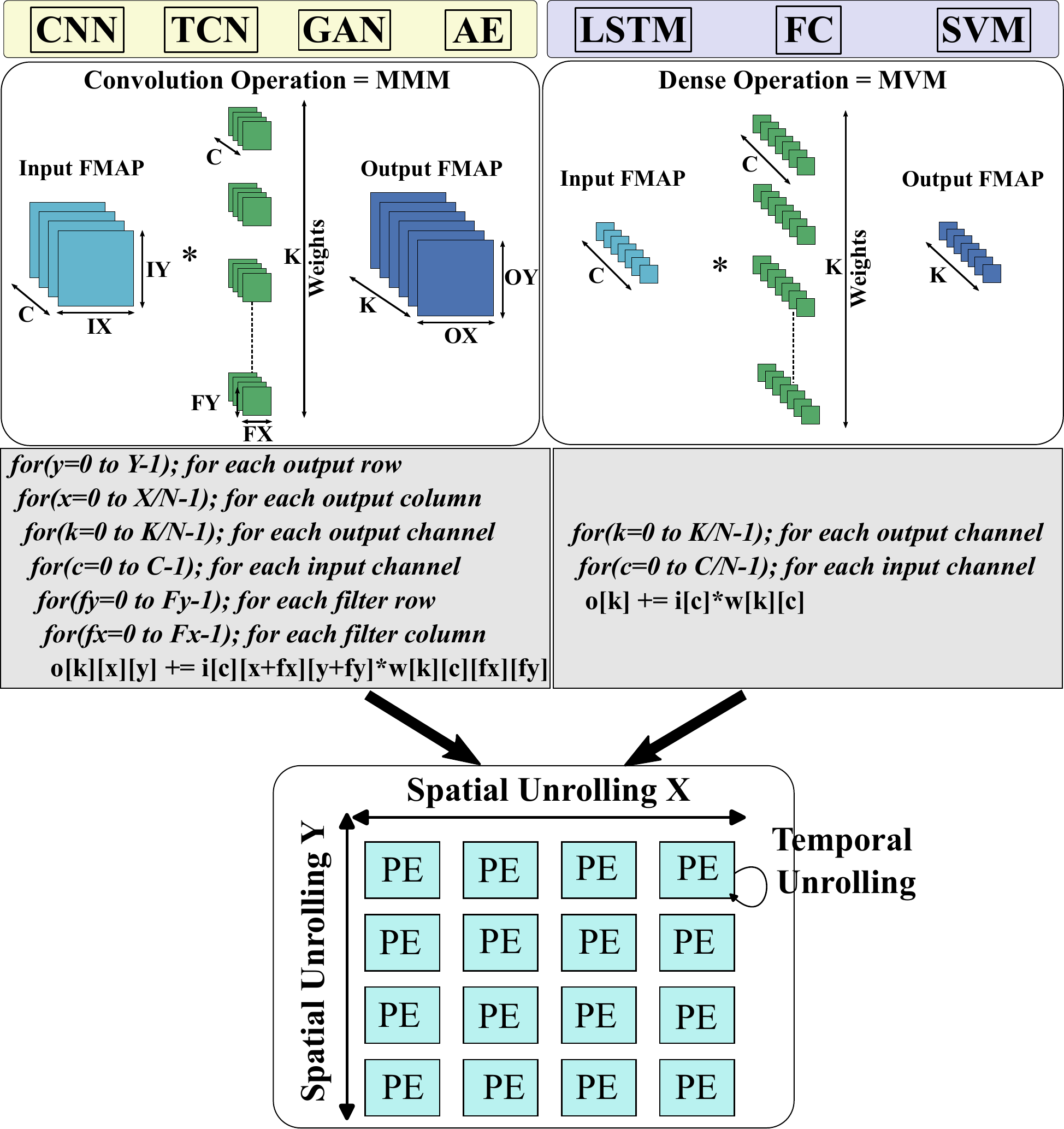}
\caption{Different ML models and their mathematical representation in terms of MMM and MVM. The nested \textit{for} loop representation can be mapped onto specialized accelerators through spatial and temporal unrolling.}
\label{fig_workloads}
\end{figure}

Convolutional and dense layers are the most common compute kernels used in DNNs and they can be decomposed into matrix-matrix multiplication (MMM) and matrix-vector multiplication (MVM) resp.. These two matrix operations can be represented mathematically as nested \textit{for} loops as shown in Fig.~\ref{fig_workloads}. Most ML compute kernels can be categorized into one of these two mathematical operations, with some special layers requiring extra hardware changes. One such kernel is the TCN layer which can be represented as a 1D CNN and requires extra support for programmable dilation which is similar to strides in a convolution. 
Recurrent neural networks (RNN) like long short-term memory (LSTM) and gated recurrent unit (GRU) can be decomposed to MVM with need for extra hardware for activation functions. These hardware changes would be discussed further in Section~\ref{sec:accel}. 

When mapping MMMs and MVMs on specialized hardware accelerators, the nested \textit{for} loops can be unrolled spatially and temporally, which is called dataflow in literature~\cite{zz}. On a 2D processing element (PE) array, two \textit{for} loops can be spatially unrolled, i.e., the loops can be parallelized along the X and Y dimensions, as shown in Fig.~\ref{fig_workloads}. In the rest of the paper, this spatial unrolling is represented as (Spatial Unrolling X)$|$(Spatial Unrolling Y). The remaining \textit{for} loops are temporally unrolled, i.e., sequential execution. Depending on the available parallelism and available re-usability, the spatial unrolling (X and Y) needs to be configurable, to be able to efficiently map all workloads, detailed in Section~\ref{sec:df_recon}.

\subsection{Deconvolution}
Autoencoders used in many machine monitoring applications consist of an encoder and a decoder pair, which tries to reconstruct the input data. After training on normal data, a reconstruction error signals 
an anomaly in the test data. 
Deconvolution or transposed convolution are used in these autoencoders and are built by combining the convolution and upsampling into a single operation. Deconvolution can be mapped as a convolution (MMM) but needs extra hardware to support zero-skipping of input for efficient mapping. 
Hardware modification can improve the mapping efficiency of this operation, and better exploit its inherent sparsity, as will be discussed in Section~\ref{sec:zero_skip}. 

\subsection{Support Vector Machines (SVMs)}
SVMs are ML algorithms used for classification and regression tasks. When classification of input data between normal behavior and an anomaly is required, a binary classifier called a one-class support vector machine (OC-SVM) can be used~\cite{ocsvm_1, ocsvm_2}. 
The decision function of a OC-SVM using the radial basis function (RBF) kernel is given by the equation~\eqref{svm_normal}. For the Laplacian kernel, the L2 norm is replaced by L1 norm. 

\begin{equation}
\label{svm_normal}
f(x) = \sum_{i=0}^{N} \alpha_{i} \cdot {\exp^{\frac{{-\lVert x-sv_{i} \rVert}_2}{2{\sigma}^2}}} - b
\end{equation}

where $x$ is the input vector with length $D$, $sv$ are the support vectors with length $D$, $N$ is the number of support vectors, $\sigma$ the standard deviation, $\alpha$ the Lagrange multiplier, and $b$ the bias. The number of support vectors $N$, in combination with the vector length $D$, can become large in these workloads, making the L1 and L2 norm calculation complex, and their deployment can gain orders of magnitude in performance when deployed on specialized accelerators. 
The $D$ and $N$ dimensions of the norm operations can be treated similar to $C$ and $K$ dimensions of a dense layer (MVM) and can be spatially unrolled on the PE array. 
In addition to unrolling the norms, extra hardware to support squaring, subtraction, rounding and absolute operation needs to be added to each PE. 
The result of the norm calculation can then be used by a CPU core to compute the overall kernel. 

\subsection{Structured Sparsity}
Exploiting sparsity in DNNs can help to reduce the computational complexity and memory requirements, by skipping zeros and compressing the NN parameters. However, random pruning or unstructured sparsity tends to be hard to efficiently map on hardware and requires special logic for zero-skipping and load balancing~\cite{han, torsten, hcgs}. The structure of sparsity (granularity of pruning) has high impact on hardware efficiency and prediction accuracy. Some works have found that unstructured sparsity achieves better prediction accuracy than structured sparsity but structured sparsity tends to be more hardware amenable and improves computational efficiency~\cite{torsten}. Thus, a structured sparse model could be trained with more iterations to revert back closer to the same prediction accuracy achieving similar overall efficiency/cost. Moreover, more coarse-grained sparsity can reduce the additional memory requirements imposed for storing indices of non-sparse data.  

With all of these diverse ML workloads and their characteristics in mind, a platform which can efficiently map all of the above, needs to be designed.

\section{TinyVers hardware architecture}
\label{sec:tinyvers}
\begin{figure}[!t]
\centering
\includegraphics[width=\columnwidth]{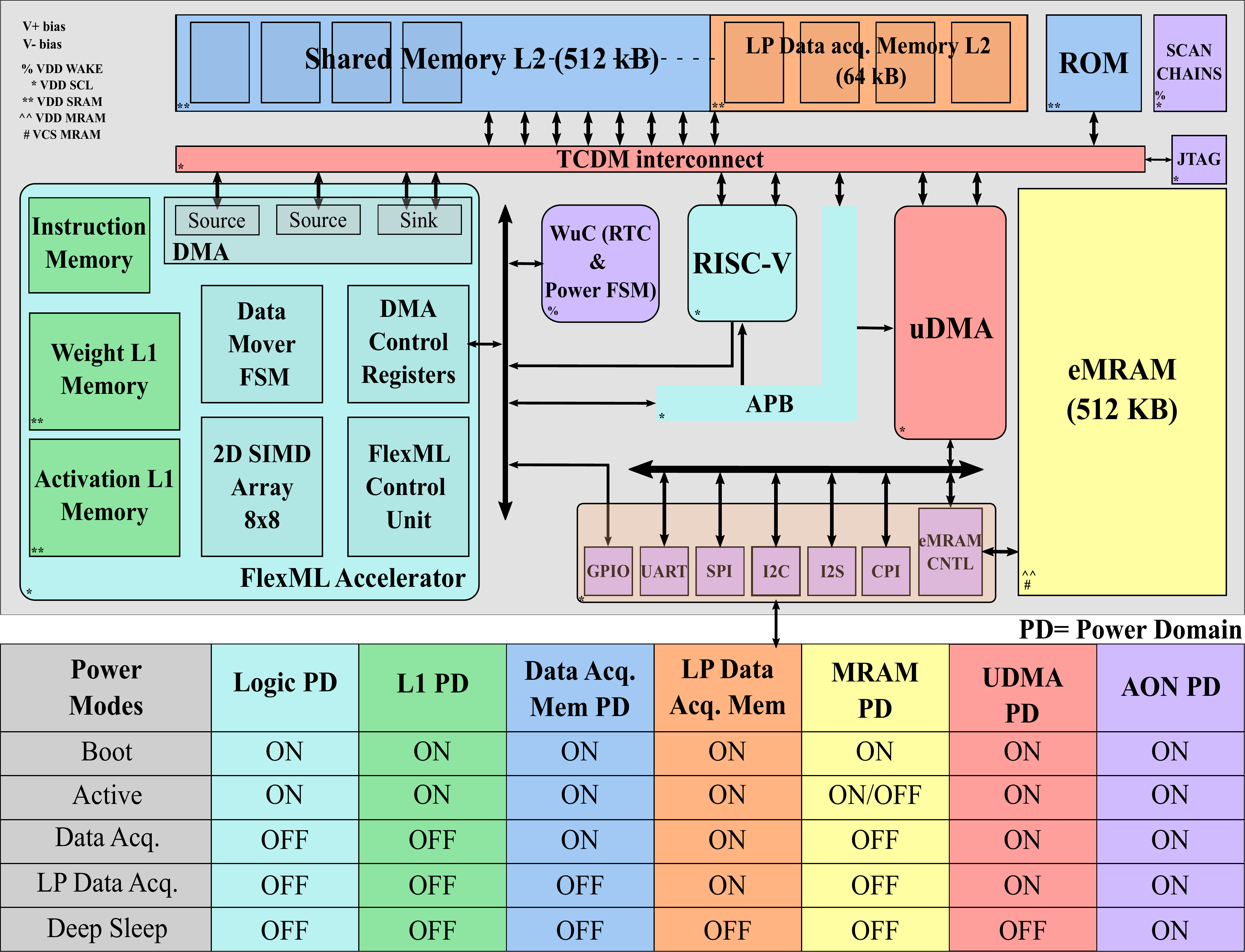}
\caption{Overview of the complete TinyVers SoC showing the different power domains (PD) with their constituting modules and the power modes supported.}
\label{fig_1}
\end{figure}

TinyVers, as shown in Fig.~\ref{fig_1}, is a heterogeneous SoC consisting of a single core RISC-V processor, a flexible ML accelerator called \textit{FlexML}, a 512 kB shared level-2 (L2) SRAM memory, a micro-DMA (uDMA) for data movement between peripherals/memory, a 512 kB eMRAM for non-volatile storage, and a WuC for power management. The SoC development is rooted in the PULPissimo platform~\cite{pulpissimo}. It embeds a 2 kB read-only memory (ROM), which acts as the first stage boot loader (FSBL) and also controls boot from JTAG, external SPI flash or the eMRAM. Two communication busses are used: 1.) a logarithmic interconnect, which enables a tightly-coupled data memory (TCDM) providing single cycle access to the shared L2, and 2.) the APB standard bus, which is used for controlling different memory mapped modules. 

The interface between the SoC and FlexML accelerator is based on the HWPE framework presented in~\cite{conti}. Using the streamers from~\cite{conti}, data is moved to-and-from the shared L2 memory with the help of FlexML's DMA engine which is a FSM controlling the data (un)loading of its private memories and double buffering operation. Several peripheral interface protocols are supported by the SoC including UART, SPI, I2C, I2S, and CPI, in addition to having 32 general purpose IOs (GPIO). Separate clocks are used for the main core logic, the peripheral interfaces, and the always-on domain which includes the WuC and the IO pads. 



\subsection{Smart Sensing Modes for TinyML}
\label{sec:smartsense}

IoT tinyML applications typically operate by collecting data across a specified time window through an array of sensors, after which the collected data can be processed to make decisions. In many applications, the time window across which the data needs to be collected before processing can start, can vary from a few $ms$ to $sec$. Moreover, during the sensor data collection, many modules of the MCU are not used since no heavy processing is done yet. This brings opportunities in improving power saving in many tinyML applications: during data collection, only the modules necessary for moving the windowed data from the sensor peripheral interfaces to the memory need to remain active, while e.g. the CPU can be put to sleep. Furthermore, in applications which work on time series data like audio, the memory requirement for the windowed data is small ($<64$ kB), such that also a large part of the main memory of the MCU can be powered-down to avoid leakage power of the unused memory section.

\begin{figure}[!t]
\centering
\includegraphics[width=\columnwidth]{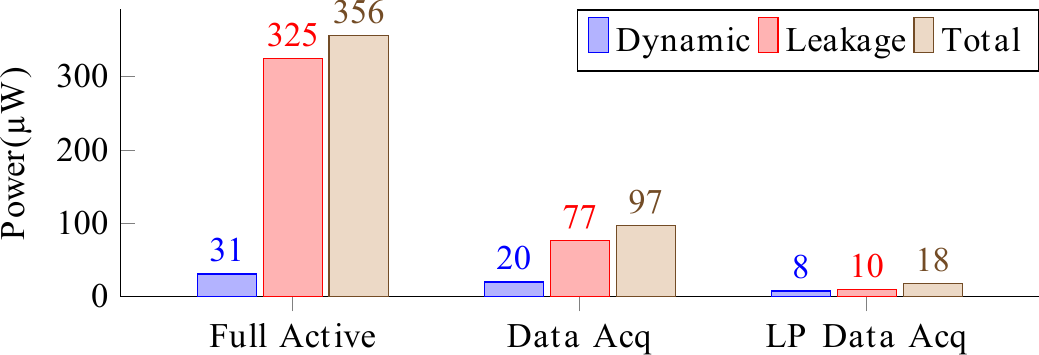}
\caption{Power simulation of post-synthesis netlist undertaken in Cadence Genus tool for the three power modes. In all the three modes, I2S data is collected at a sampling frequency of 44.1 kHz for a window of 2 seconds. Full active power reported includes configuration of uDMA by RISC-V core and interrupt handling procedure, in addition to data collection.}
\label{fig_3}
\end{figure}

To this end, TinyVers introduces two tinyML optimized data acquisition power modes: 1.) `Data acq.' and 2.) `LP data acq.', as shown in Fig.~\ref{fig_1}. The data acq. mode, targeted towards applications with large sample data like vision, keeps the uDMA module and the complete shared L2 memory (512 kB) powered up. In contrast to that, the LP data acq. mode only keeps part of the shared L2 memory (64 kB) powered up, in addition to the uDMA. This mode is specifically targeted towards applications which needs time series and audio data like keyword spotting, machine monitoring, biosignal analysis, etc. Fig.~\ref{fig_3} shows an estimation of the power saving that can be achieved when moving from a full active mode to the two tinyML sensing modes, with almost 3.5$\times$ improvement between the full active and data acq. modes and 5.5$\times$ between data acq. and LP data acq. modes.  

\subsection{Power Management}
\label{sec:pm}

Aggressive power management is pursued in TinyVers on top of standard low power design. The SoC is divided into 6 switchable power domains and 1 always-on domain (AON), as shown in Fig.~\ref{fig_1}. Each switchable power domain consists of multiple power gating switches, which isolate the VDD of the power domain from the global VDD supply. These power gating switches are controlled by control signals driven from the WuC of the AON domain. All interconnect crossings between the power domains are equipped with bidirectional level shifters and isolation cells, such that the individual supply voltages of the domains can be controlled independently.

\begin{figure}[!t]
\centering
\includegraphics[width=\columnwidth]{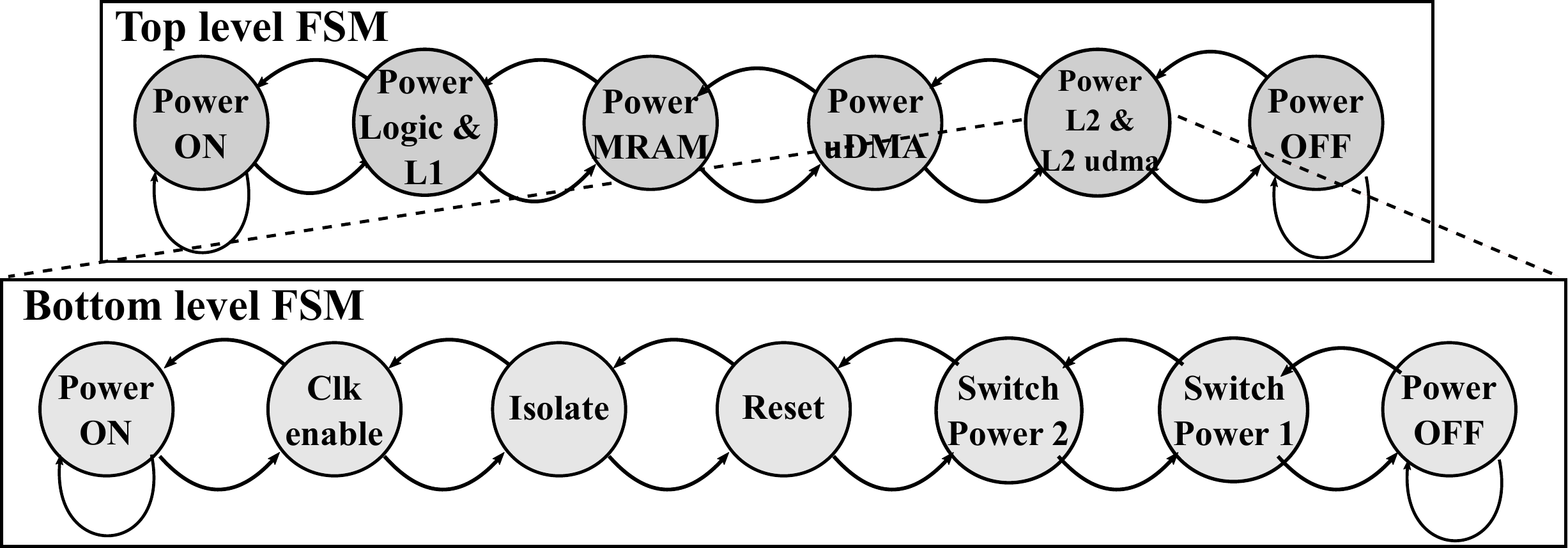}
\caption{Flow diagram showing the hierarchical FSM used in the WuC.}
\label{fig_2}
\end{figure}

The smart WuC is in charge of this power management control, relying on 
a real-time counter (RTC). The counter can be programmed by the RISC-V core with millisecond granularity. The RISC-V core can instruct the WuC to bring the SoC into one of the five supported power modes shown in Fig.~\ref{fig_1}. To this end, the WuC encompasses hierarchical finite-state machines (FSM) driven by the RTC, as shown in Fig.~\ref{fig_2}, controlling the power-up and power-down of the complete SoC and the different power domains. The top level FSM controls the sequence of power-up/down of the different power domains and the bottom level FSMs control the fine-grain sequence to (de)activate the isolation cells and the power gating switches of the individual power domains. 
Emerging memories like ReRAM, MRAM, FeRAM, PCM, etc.~\cite{emerg_1, emerg_2}, have shown promise in building cost-effective embedded non-volatile memories (NVM) targeting 
applications in edge computing for automotive or industry 4.0. NVM memories can be used as the storage space for boot code and other parameters that need to be stored. This enables two things: 1.) Duty-cycling can be used as a means of reducing power consumption in applications which do not require always-on operation; and 2.) the SoC does not need to go to a central cloud server in order to fetch its boot codes and NN parameters when it is power-cycled. Moreover, the availability of 
the NVM embedded on-chip, avoids the high energy cost of fetching data from off-chip.

MRAM promoted as a universal memory, uses magnetic polarity to store data in its bitcells~\cite{mram}. Being non-volatile and almost as dense as traditional SRAM, they are a good fit for tinyML applications using extreme edge SoCs. With this in mind, TinyVers integrates a 512 kB embedded MRAM on-chip, enabling extreme power management strategies for smart sensing and on-demand computation. In the SoC, the eMRAM acts as a non-volatile storage for the boot code that the RISC-V needs to wake-up and start processing, and can also store the NN parameters of the mapped ML workloads. 
The eMRAM can, finally, also be used as a non-volatile scratchpad space for storing windowed data in smart sensing applications. The interface between eMRAM and the shared L2 memory uses the uDMA unit and the design is based on the work of~\cite{vega}.


\section{FlexML Accelerator}
\label{sec:accel}
This section firstly describes the architecture overview of the FlexML accelerator, followed by the dataflow reconfiguration used for flexible mapping, efficient zero-skipping used for deconvolution and structured sparsity, and finally the hardware for supporting SVM, as briefly discussed in Section~\ref{sec:background}.

\subsection{FlexML Architecture Overview}
\label{sec:accel_ov}

\begin{figure}[!t]
\centering
\includegraphics[width=0.9\columnwidth]{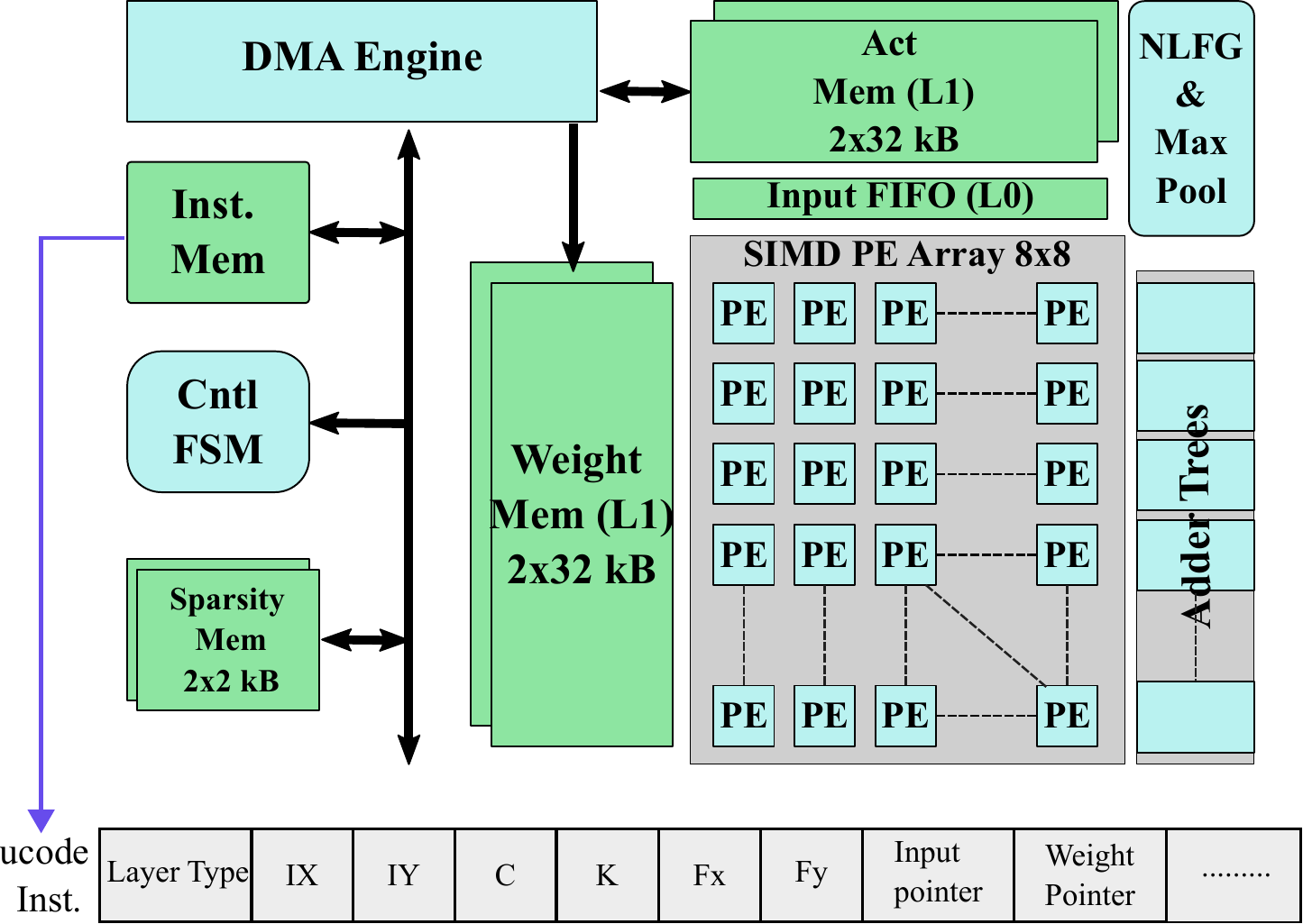}
\caption{FlexML accelerator architecture overview with ucode instruction.}
\label{fig_4}
\end{figure}

\begin{figure}[!t]
\centering
\includegraphics[width=\columnwidth]{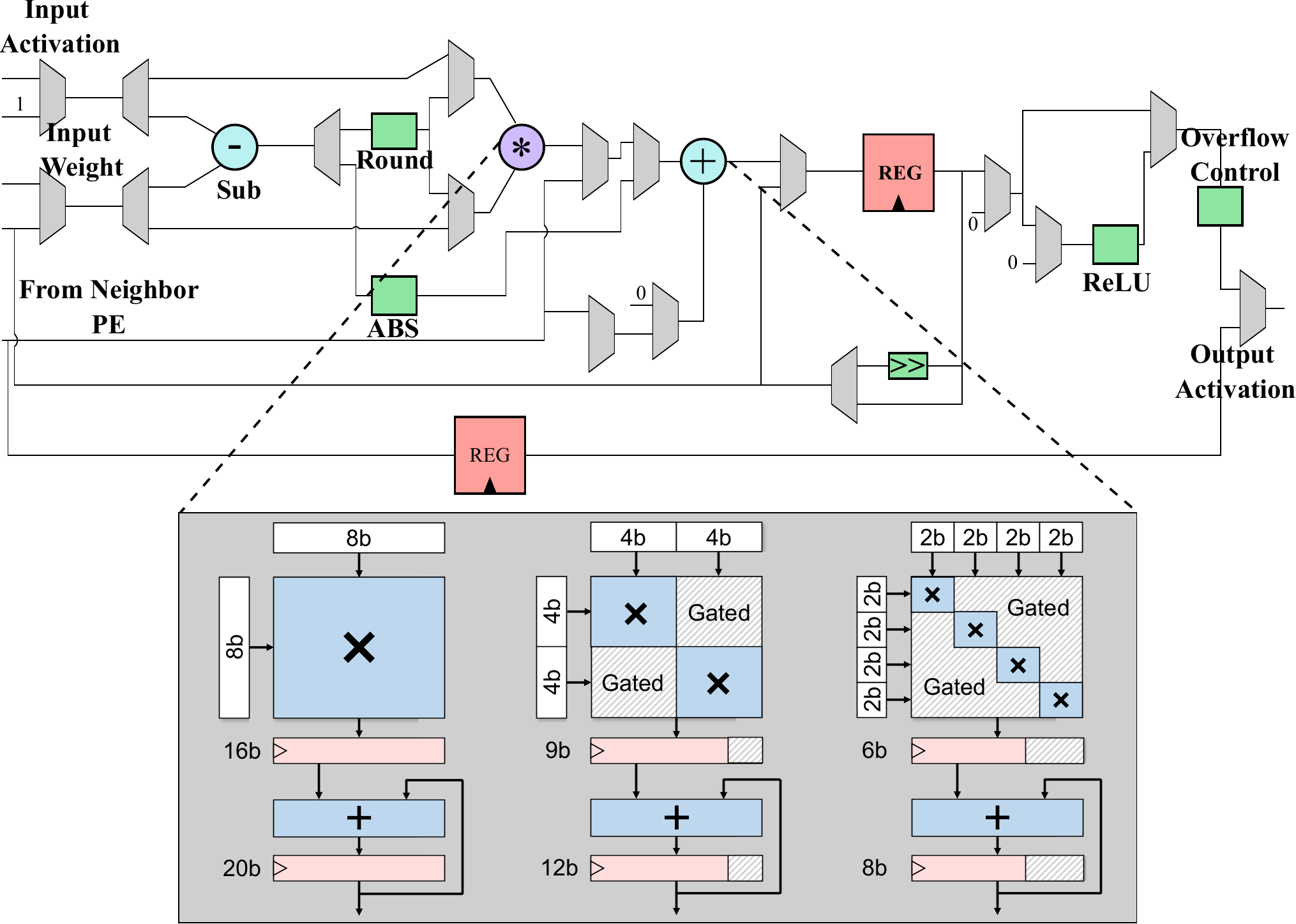}
\caption{Block diagram of the processing elements used in the flexML accelerator, showing the precision-scalable MAC unit and the additional hardware to support SVM. }
\label{fig_pe}
\end{figure}

The FlexML accelerator is TinyVers' specialized, versatile hardware accelerator. FlexML is designed to efficiently support the large diversity in ML workloads for tinyML applications, while exploiting the data reuse present in individual layer characteristics. This is achieved through a zero-latency runtime dataflow reconfiguration, discussed in Section~\ref{sec:df_recon}. As shown in Fig.~\ref{fig_4}, FlexML encompasses an 8$\times$8 single instruction multiple data (SIMD) array of processing elements (PE), wherein each processing element consists of a precision-scalable multiply-accumulate (MAC) unit with support for INT 8/4/2~\cite{mei}, shown in Fig.~\ref{fig_pe}. As a result of the precision-scalability, the SIMD array can be reconfigured to be a 8$\times$8/16/32 array of INT8/4/2 MAC units, resp.. Each PE performs 1/2/4 MAC operations per cycle based on the selected precision (INT8/4/2) and the results are accumulated in a 32-bit register with full/partial output stationarity, reducing the movement cost of the large bit-width partial sums. The final output is passed through a ReLU function (if enabled)
, followed by re-quantization to the selected precision and written back to the activation L1. Mixed precision quantization can help in improving performance of DNN models when moving below 8 bits precision. However, the hardware overhead of mixed precision can reduce the overall efficiency of PEs due to varying bandwidth and serialized dataflow~\cite{vincent}. Thus, FlexML only supports symmetric precision for its weights and activation. In addition, a simple shift and ReLU is used for normalization of output, which also keeps hardware overhead low. In order to maintain accuracy of the models, a hardware aware training framework, mentioned in Section~\ref{sec:compile}, is used.

Supporting the SIMD PE array, are private level-1 (L1) SRAM based memories for storing both weights (64 kB) and activations (64 kB). Both the weight L1 and activation L1 are composed of two 32 kB banks operating in a ping-pong manner to overlap data writing and reading, improving the overall performance. 
An intermediate memory level L0 is provided between the activation L1 and the PE array. This L0 memory is a FIFO buffer of size 16$\times$8 bits, used to improve data locality when doing shifting window operation in convolution. Furthermore, a separate non-linear function generator (NLFG) and a max pooling unit are provided. The NLFG uses LUT-based linear approximation to generate the various activation functions (other than ReLU) used in NN models such as tanh, sigmoid, etc. To control the dataflow and control flow inside the accelerator, a control unit with FSMs fetches ucode instructions from the instruction memory, decodes the instruction and deploys the relevant layer on the PE array by updating the control signals and counters that track the workload. The ucode instructions are generated by a pseudo-compiler built in python (Section~\ref{sec:compile}), and consists of CISC-like layerwise long instructions with hyperparameters and shown in Fig.~\ref{fig_4}. The control unit is also extended to enable support for efficient zero-skipping of activations in the case of deconvolution and zero-skipping of pruned weights in conjunction with the sparsity index memories (Section~\ref{sec:zero_skip}).


\subsection{Dataflow Reconfiguration}
\label{sec:df_recon}

\begin{figure}[!t]
\centering
\includegraphics[width=\columnwidth]{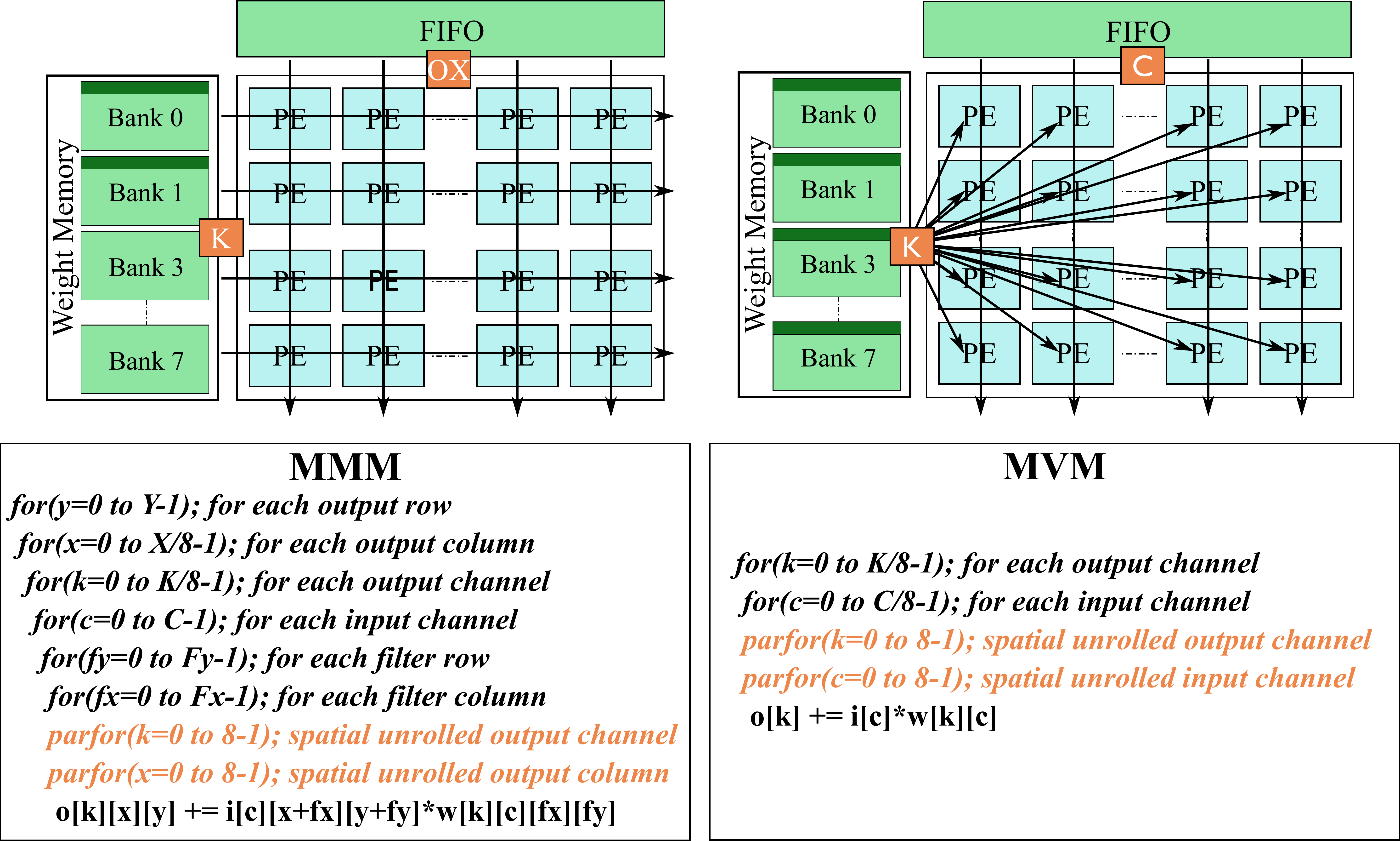}
\caption{Diagram showing the dataflow reconfiguration used to switch from OX$|$K dataflow (left) for MMM to C$|$K dataflow for MVM. The nested \textit{for} loops below show the addition of parfor loops for the spatial unrolling used.}
\label{fig_5}
\end{figure}

In order to efficiently map the diverse set of ML workloads, runtime dataflow reconfiguration is supported in the FlexML accelerator at no latency overhead. The configurability enables efficient mapping of both: 1.) MMMs used for CNN, deconvolution and TCN, exploiting both input and weight spatial data reuse under an OX$|$K dataflow with output stationarity, and 2.) MVMs used for FC, RNNs and norm calculation of SVMs with batch size 1, exploiting the available input spatial data reuse under a C$|$K dataflow with partial output stationarity. Multiple previous works have proposed dataflow reconfiguration in hardware to optimally map different workloads~\cite{flexflow,maeri,eyerissv2}. However, these works suffer from large hardware overhead and latency for diverse dataflow support and are not suitable for extreme edge devices. This work limits the dataflows to two optimal mapping schemes, thereby, keeping hardware and power overhead low. Moreover, none of the prior works have looked into mapping of TCN, AE, and SVM on the same hardware accelerator. Fig.~\ref{fig_5} shows the OX$|$K (left) and C$|$K dataflow (right) and their hardware implementation, resp.. 
In the OX$|$K dataflow, the spatial unrolling is applied to the OX and the K dimension of the nested \textit{for} loop, allocating the unrolled OX dimension along the columns and the unrolled K along the rows of the SIMD PE array of dimension 8$\times$8. 

The rest of the for loops are temporally unrolled as shown in the nested \textit{for} loops in Fig.~\ref{fig_5}, resulting in an output stationary dataflow. 
Under this dataflow regime, the activation L1 memory multicasts input activation data in the vertical dimension to the L0 FIFO memory, which fetches 8 words in the first cycle followed by single word during the shifting window operation, thereby, reducing the memory bandwidth and number of memory fetches by utilizing the reuse opportunity. The weight L1 memory provides data in the horizontal dimension, providing 8 words using 2 internal banks, where each word is multi-cast along the row. Due to the output stationarity, accumulation continues till the final output is generated which are then systolically shifted out vertically to the activation L1, requiring 8 cycles to complete the output write-back. The input data shifting inside the L0 FIFO 
is made programmable to support the variable dilation used in TCNs or variable strides in general.

The alternative C$|$K dataflow is used for MVM, as this workload cannot utilize the OX$|$K dataflow efficiently due to lack of re-usability of weights. 
Under this dataflow, the C dimension is spatially unrolled along the vertical column dimension and the K dimension along the horizontal row dimension. The activation L1 memory multicasts 8 words of input activation along the vertical dimension, bypassing the L0 FIFO memory. With a batch size of 1, no weight reuse is available and, thus, each PE needs a new  weight every cycle. In order to meet this requirement, the weight memory utilizes all of its 8 banks to unicast 64 different weight words to the PEs. PE rows operate on different input channels ($C$) of the same output channel ($K$). 
Hence, once the required MAC operations per PE are done, the outputs of PEs of the same row are accumulated using an adder tree and one final output per row is shifted out to the activation memory. 



\subsection{Efficient Zero-skipping for Deconvolution and Blockwise Structured Sparsity}
\label{sec:zero_skip}

\begin{figure}[!t]
\centering
\includegraphics[width=\columnwidth]{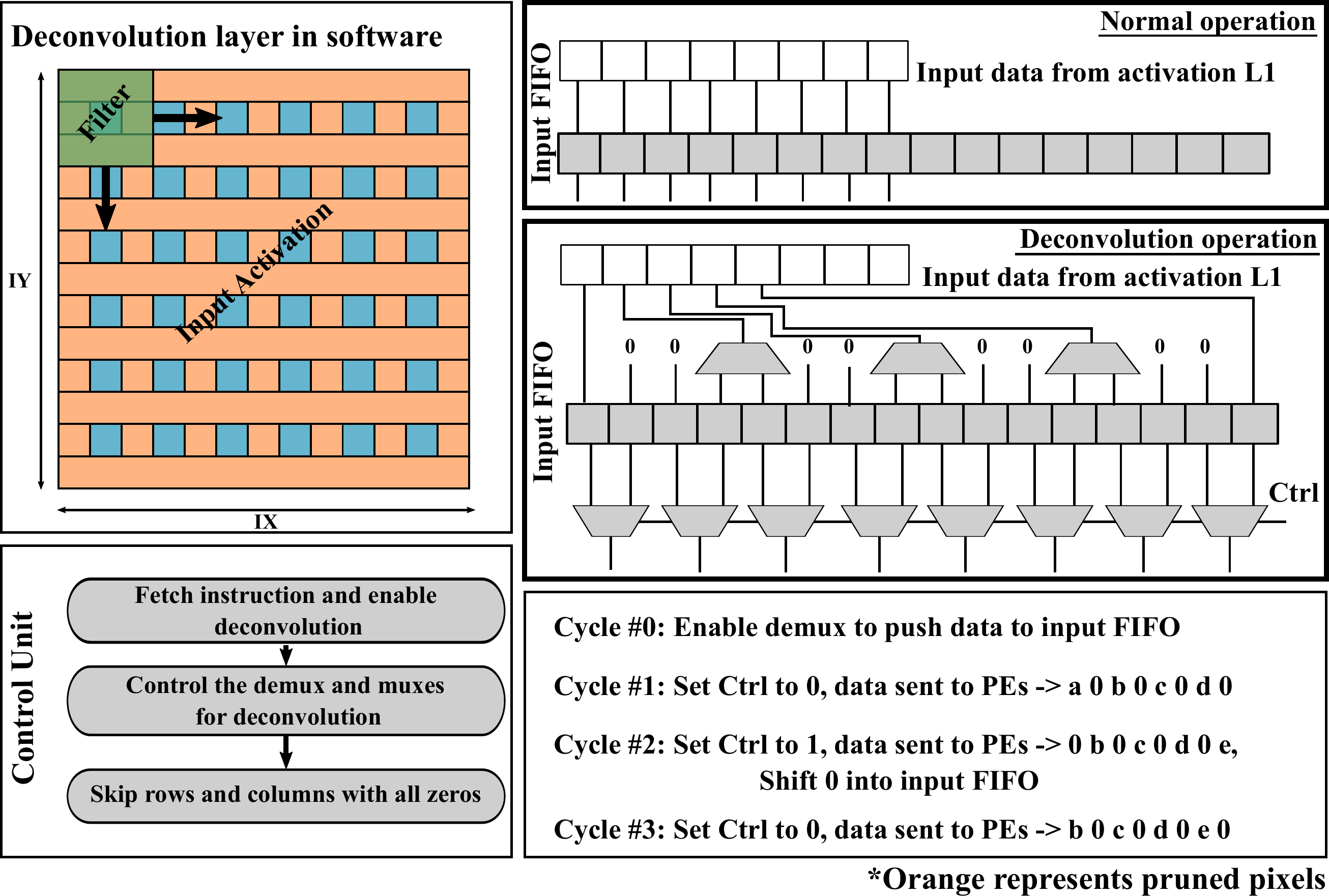}
\caption{Representation of deconvolution layer in software (top left), control unit running the zero-skip operation (bottom left), the architectural change required on the L0 FIFO to support deconvolution (top right), and cycle by cycle operation of the FIFO and PEs (bottom right).}
\label{fig_deconv}
\end{figure}


The FlexML accelerator supports efficient zero-skipping of deconvolution workloads. As shown in Fig.~\ref{fig_deconv}, the input FIFO is designed such that when in deconvolution mode, it only fetches one set of words and shuffles it with zero padding. 
The control unit skips the rows and columns with zeros that would result in redundant computation, resulting in 
a performance gain of up to 2$\times$ compared to running deconvolution in convolution mode with upsampling.

\begin{figure}[!t]
\centering
\includegraphics[width=0.89\columnwidth]{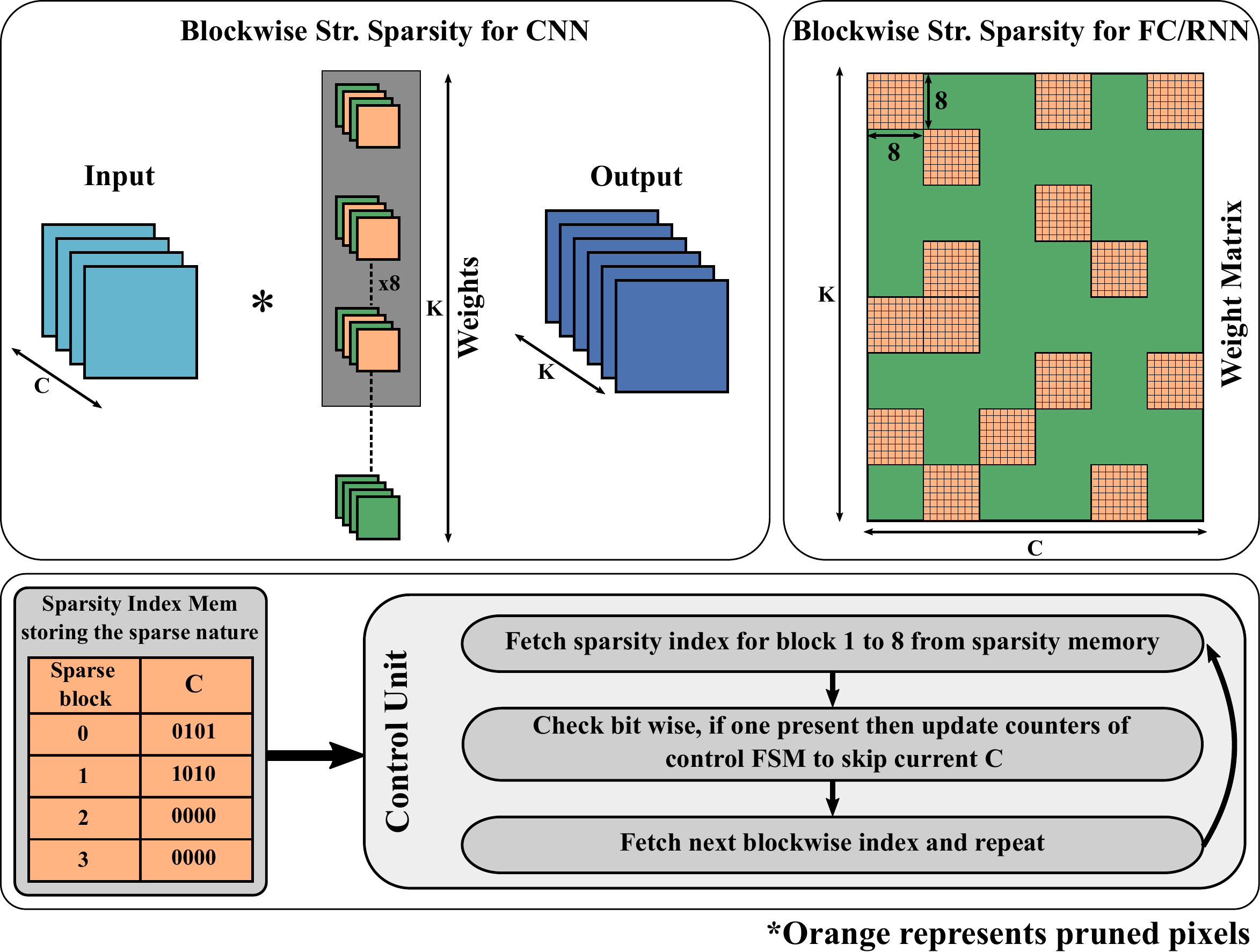}
\caption{Blockwise structured sparsity applied to CNN and dense layers (top), control unit operation in tandem with sparsity index memory to support zero-skipping (bottom).}
\label{fig_bss}
\end{figure}

TinyVers also supports structured sparsity, more specifically, blockwise kernel-level sparsity (2D) for both convolutional and dense layers~\cite{han, hcgs}. In this scheme, shown in Fig.~\ref{fig_bss}, complete input channels of the filter kernels are pruned with a constraint 
that a block size of 8 filter kernels ($K=8$) should share the same pruning. The block size is decided by the dimension of the PE array and the spatial unrolling of $K$ along the horizontal dimension of the 2D PE array. 
In our case, the selected block size makes controlling the dataflow and control flow easier. Applying the same channel pruning to all the 8 filter kernels mapped in parallel on the PE array makes the mapping efficiency higher as all the rows can still operate with a common control logic, and enables not only energy savings, but also throughput benefits. For this, the FlexML accelerator consists of specialized sparsity index memories which store the bit encoded indices of the pruned channel groups. 
Fig.~\ref{fig_bss} shows the sparsity index memory and the control flow logic used in the control unit. Before every filter kernel block increment, the control unit fetches an index memory word and checks the data bit-by-bit for sparsity state, as the input channels increment. If a sparse channel is detected, the complete computation of the channel is skipped, thus, avoiding any zero computation.


\subsection{Support Vector Machine}
\label{sec:svm}

The L1 and L2 norm of OC-SVM requires modification of the PEs in order to use the same hardware for mapping the workload. 
As shown in Fig,~\ref{fig_pe}, each PE is extended with a subtraction block, absolute unit, rounding unit, and the modification of the multiplier to also enable squaring for the norm calculation within the PE array. The input data vector $x$ and the support vector $sv_{i}$ are of dimension $D$ 
and the number of support vectors is $N$. 
When used in the C$|$K dataflow, the $D$ dimension of the input data vector of $x$ is unrolled and multicasted vertically (C) along the PE array, 
while the $N$ dimension of the support vector $sv_{i}$ are unrolled and unicasted horizontally (K). The results of the $N$ norm calculations, computed in the PEs, are then sent to the shared L2 memory where it is then post-processed by the RISC-V core with the GNU C in-built exponential function, multiplication with $\alpha$ and summation over $N$ to generate the final output shown in equation~\eqref{svm_normal}.  


\section{Deployment of neural networks on TinyVers}
\label{sec:compile}
Hardware used for ML applications also requires a user programmable full stack that can translate ML algorithms directly from existing ML training and inference frameworks like Tensorflow, Keras, PyTorch, etc. This makes the quick and easy deployment of various ML workloads onto an existing hardware possible. 
A python based pseudo-compiler framework created for TinyVers taking into account its heterogeneity is created. An ML algorithm 
is first quantized to selected precision using the QKeras framework~\cite{qkeras} for quantized-aware training. The quantization-aware training framework takes into consideration the hardware constraints such as symmetric quantization and the shift based scaling of output in the PEs of the accelerator. The quantized model is then passed to a python-based NN compilation which takes in the hardware description and provides a set of C-based header files for the RISC-V core, consisting of ucode instructions for the accelerator, NN parameters and also a golden model for verification of the mapped workload. 

\section{Chip Implementation and Measurement}
\label{sec:results}
\begin{figure}[!t]
\centering
\includegraphics[width=\columnwidth]{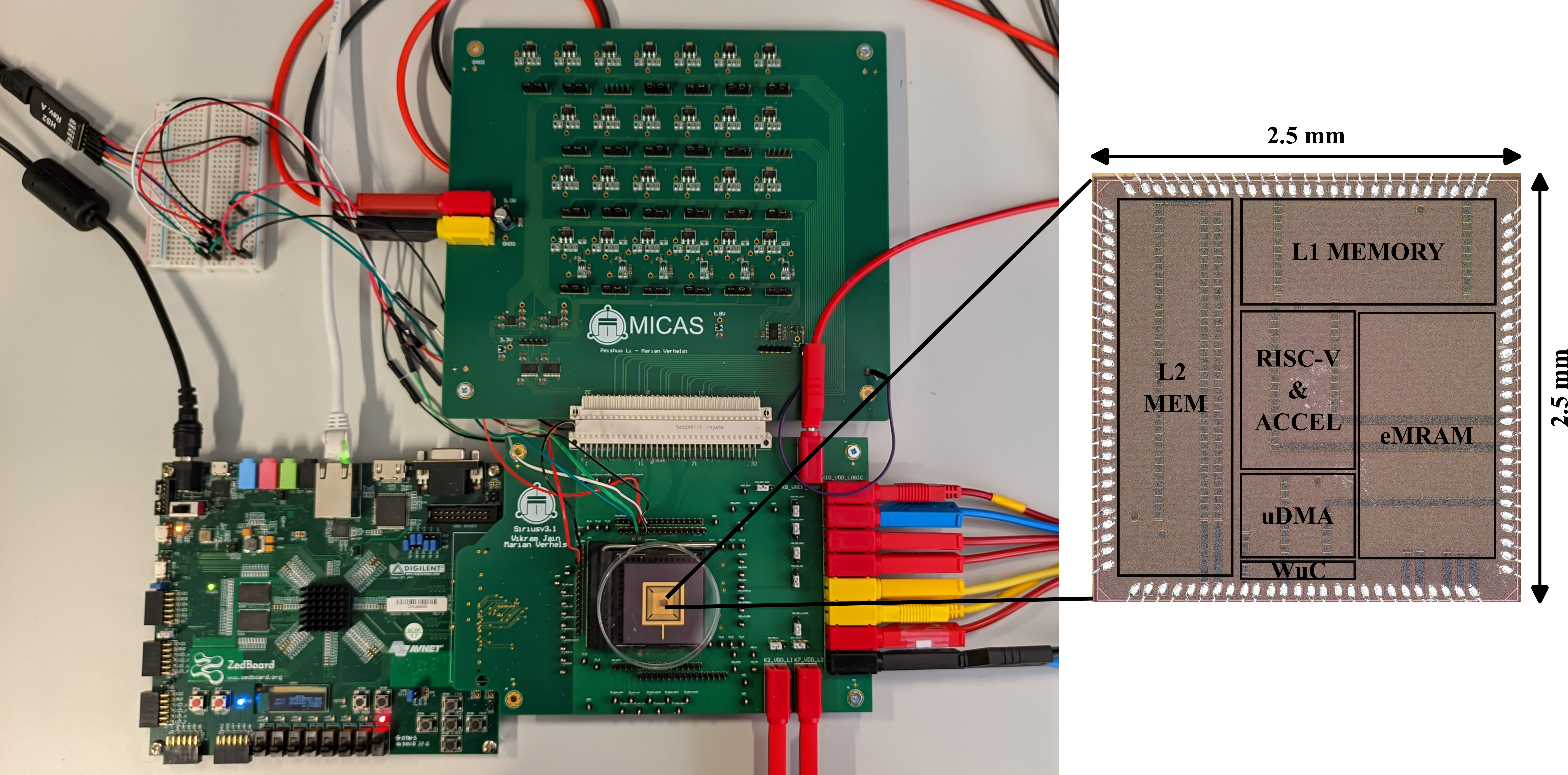}
\caption{Measurement setup and chip microphotograph.}
\label{fig_10}
\end{figure}

The TinyVers chip microphotograph shown in Fig.~\ref{fig_10} was implemented and fabricated in GlobalFoundries 22FDX\trademark{}. The figure shows the different sub-modules used in the SoC and detailed in previous sections. Fig.~\ref{fig_10} also shows the lab setup used for measurements and benchmarking. The following subsections details the measurements and benchmarking done on the SoC for power, energy efficiency and performance.



\subsection{Peak Performance Analysis}
\label{sec:ppa_cnn}

\begin{figure}[!t]
\centering
\includegraphics[width=0.9\columnwidth]{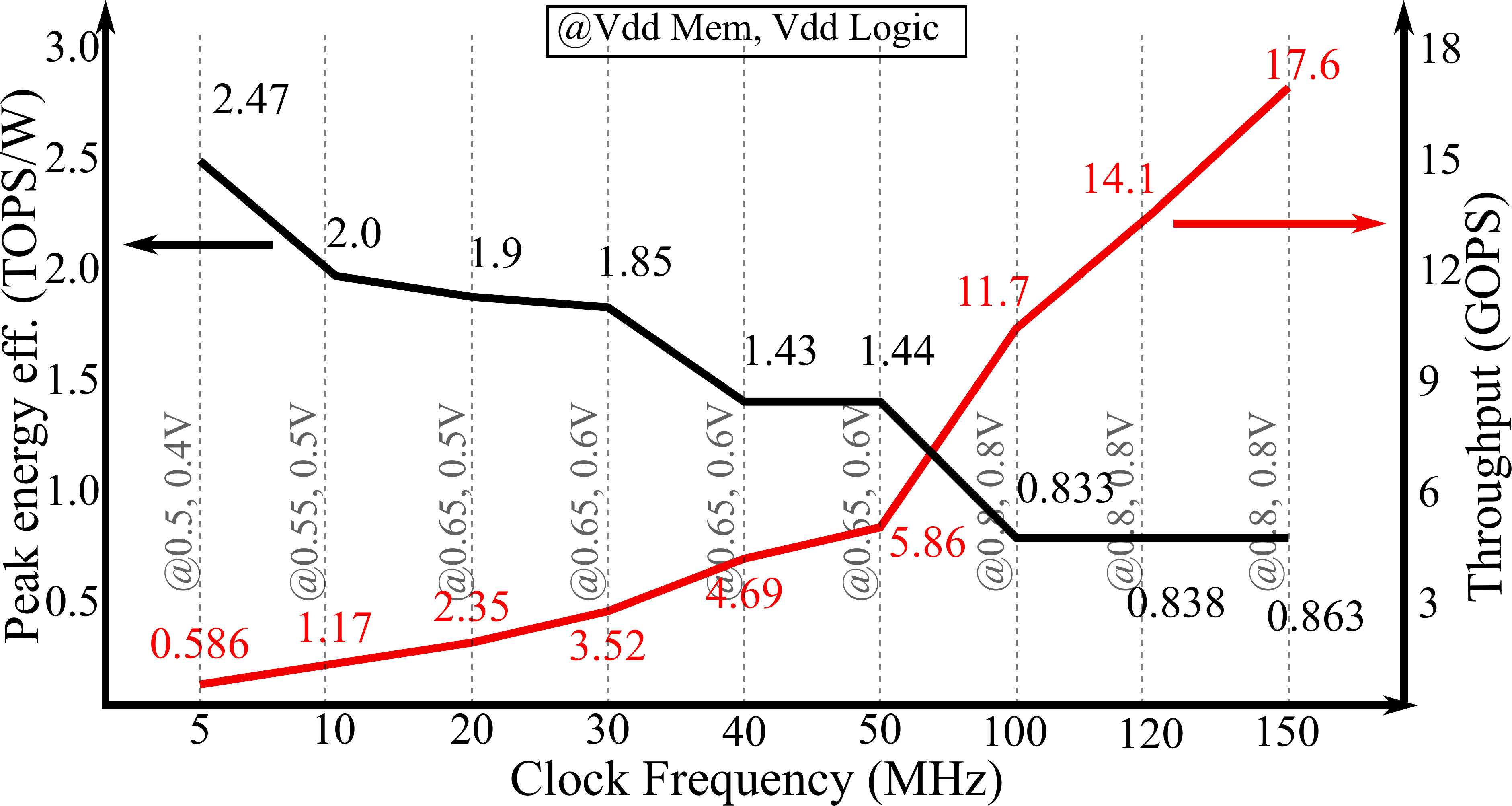}
\caption{Peak performance analysis of CNN3$\times$3 layer.}
\label{fig_11}
\end{figure}

\begin{figure}[!t]
\centering
\includegraphics[width=\columnwidth]{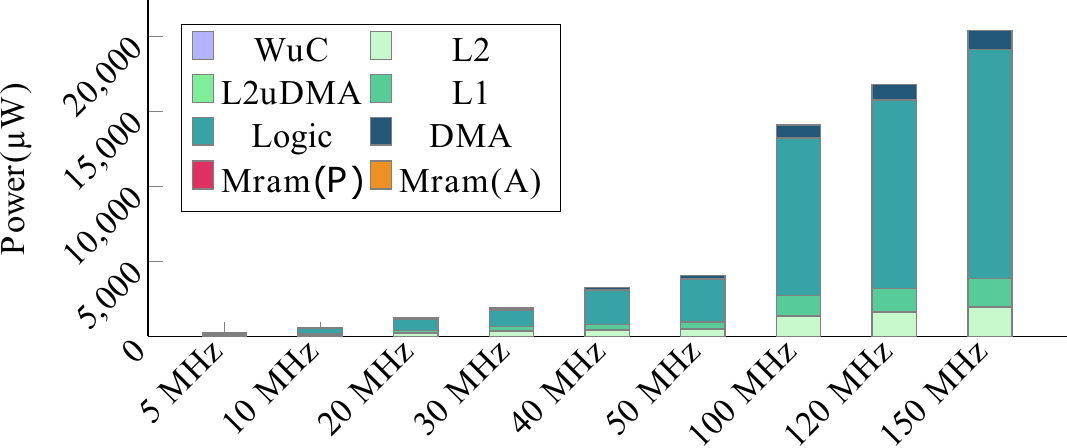}
\caption{Power breakdown of the peak perf. analysis with CNN3$\times$3. MRAM power consumption is negligible as it is OFF in active mode. MRAM(A) and MRAM(P) represents MRAM array and MRAM periphery resp..}
\label{fig_cnn_breakdown}
\end{figure}

First, a peak performance analysis is undertaken using a single CNN layer with 32 input channels, 32 output channels and a 3$\times$3 filter kernel. Selection of the used layer for peak performance is driven by the fact that convolutional layers with a 3$\times$3 filter kernel are the most commonly used layer in modern DNN models. The hyperparameter selection of the CNN layer is driven by the constraint of maximum utilization of the PE array and the size of the private L1 memories of the accelerator. The 8 bit quantized activation and non-sparse (structured) weights of the CNN are generated using the compiler framework using the Google speech dataset for keyword spotting~\cite{gcsd} and verified against the golden model for functional correctness.

Fig.~\ref{fig_11} plots the peak energy efficiency and the throughput with respect to the clock frequency while sweeping the voltage supply of the logic and memories for the benchmarked CNN layer. For fair comparison with other SotA chips, no body biasing is applied. Fig.~\ref{fig_cnn_breakdown} shows the power breakdown of individual modules when running the benchmarking layer. 
The SoC shows a large flexibility in delivered performance ranging from high energy efficiency/low throughput of 2.5 TOPS/W, 586 MOPS when operating at a clock frequency of 5 MHz with 0.4 V logic, 0.5 V memories, to low energy efficiency / high throughput of 0.8 TOPS/W, 17.6 GOPS operating at 150 MHz with 0.8 V logic and memories. This provides a large range for extreme edge tinyML applications to operate, trading-off between speed and energy efficiency. 


\subsection{Workload Benchmarks}
\label{sec:ppa_workloads}

\renewcommand{\arraystretch}{1}
\begin{table}[!t]
\caption{Workload benchmarks}
\label{tab:workload_details}
\centering
\begin{tabular}{c|ccccc}
\toprule
 & \textbf{Workload} & \begin{tabular}[c]{@{}c@{}}\textbf{Acc.} \end{tabular} & \begin{tabular}[c]{@{}c@{}}\textbf{Power}\\ \textbf{($\mu$W)}\end{tabular} & \begin{tabular}[c]{@{}c@{}}\textbf{Peak} \\ \textbf{perf.} \\ \textbf{(GOPS)} \end{tabular} & \begin{tabular}[c]{@{}c@{}}\textbf{Peak} \\ \textbf{(effective NZ)} \\ \textbf{energy eff.} \\ \textbf{(TOPS/W)}\end{tabular} \\ \midrule
\multirow{8}{*}{\rotatebox[origin=c]{90}{\textbf{Synthetic}}}{\begin{tabular}[c]{@{}c@{}}\end{tabular}}
 & CNN@8b & - & 237  & 0.586 & 2.47(2.47) \\ \cline{2-6}
 & CNN@4b & - & 197  & 1.17 & 5.94(5.94) \\ \cline{2-6}
 & CNN@2b & - & 197  & 2.35 & 11.9(11.9) \\ \cline{2-6}
 & CNN@8b, & - & 239  & 1.03 & 4.31(2.46) \\ 
 & 50\% sparse & & & &\\ \cline{2-6}
 & CNN@8b, & - & 212  & 3.64 & 17.1(2.76) \\
 & 87.5\% sparse & & & &\\ \cline{2-6}
 & FC/RNN/SVM, & - & 140  & 0.116 & 0.829(0.829) \\
 & batch=16 & & & &\\ \cline{2-6}
 & Deconv@8b & - & 235  & 1.36 & 5.78(2.49) \\

 \midrule
\multirow{8}{*}{\rotatebox[origin=c]{90}{\textbf{\hspace{1.25cm}Real-time}}}{\begin{tabular}[c]{@{}c@{}} \end{tabular}} 
& TCN (KWS) & 93.3\%$^{*}$ & 193   & 0.204   & 1.05(1.05)\\ \cline{2-6}
& CAE   & - & 209   & 0.442  & 2.11(1.27) \\ \cline{2-6}
& ResNet-8   & 82\%$^{+}$ & 228  & 0.267  & 1.17(1.17) \\ \cline{2-6}
& OC-SVM  & - & 129  & 0.126  & 0.972(0.972) \\
\bottomrule
\multicolumn{5}{l}{$^{*}$ 12-class task, baseline=93.46\%, $^{+}$ baseline=85\%}
\end{tabular}%
\end{table}

Using the peak energy efficiency operating point (5 MHz, 0.4 V logic and 0.5 V memory) from Section~\ref{sec:ppa_cnn}, further performance analysis of different synthetic and actual real-time benchmarks are evaluated. Table~\ref{tab:workload_details} shows the SoCs flexibility through mapping of different ML layers and full workloads. The CNN layer from Section~\ref{sec:ppa_cnn} is extended and measured with different precision and blockwise structured sparsity (BSS) levels. When moving to lower precision of INT-4 and INT-2, the peak throughput improves by 2$\times$ and 4$\times$ while the peak energy efficiency improves by 2.4$\times$ and 4.8$\times$ resp., achieving a maximum of 11.9 TOPS/W at INT-2. As shown in Table~\ref{tab:workload_details}, at 8 bit precision with 50\% BSS (16/32 input channels pruned) the performance improves by around 1.7$\times$ while at 87.5\% BSS (28/32 input channels pruned) the performance increases by approximately 6.9$\times$. Further performance improvement can be gained when moving to lower precision, however, low precision combined with high BSS levels can cause a large drop in accuracy and, thus, is not explored in this benchmarking. Other synthetic benchmarks such as FC, RNN, SVM and a deconvolutional layer similar to the CNN layer in terms of hyperparameters are explored and the results are shown in the table. For the dense layers, batching of 16 is used. 

\begin{figure}[!t]
\centering
\includegraphics[width=0.9\columnwidth]{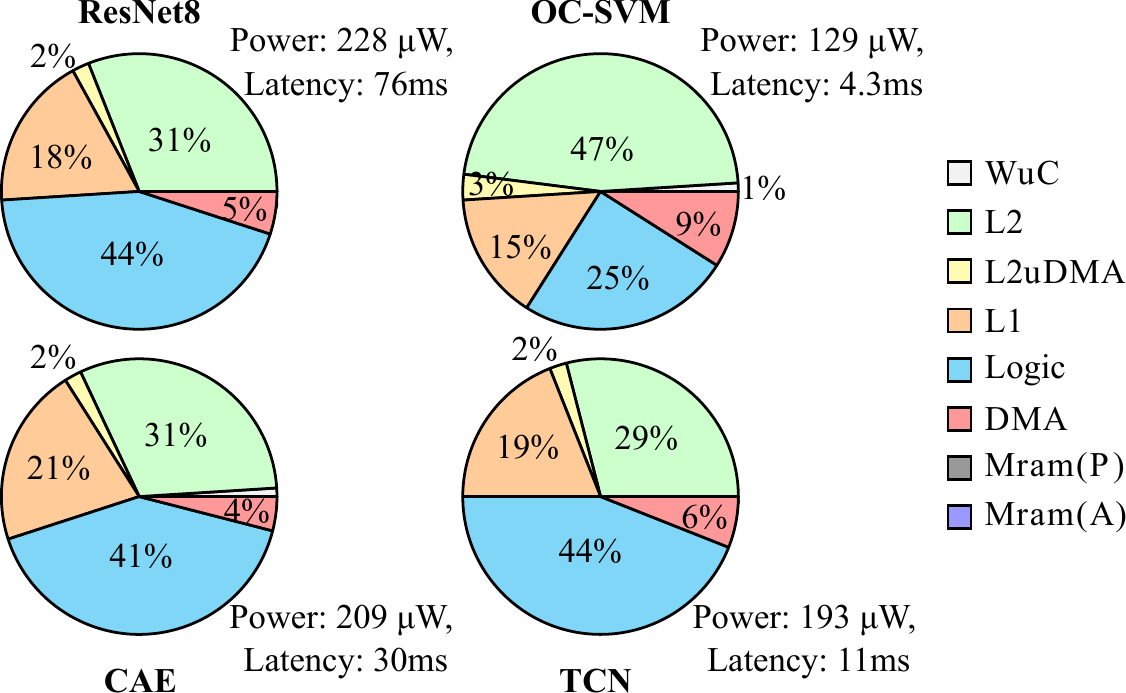}
\caption{Energy breakdown showing the distribution of measured energy of the chip modules for running a single inference of the four real-time workloads on FlexML and RISC-V with input data already available in L2 memory. The power and latency measurements starts from setting up of accelerator parameters by RISC-V, data movement from L2 to L1, inference computations, and ends with post processing by RISC-V core. MRAM power consumption is negligible as it is OFF in active mode.} 
\label{fig_pb_bmarks}
\end{figure}

Finally, 4 real-time application benchmarks are used to show the capabilities of the SoC: 1.) keyword spotting (KWS) using TCN model~\cite{tcn_work,seba} on google speech dataset, 2.) continuous machine monitoring with a convolutional auto-encoder (CAE)~\cite{cae} on MIMII dataset~\cite{mimii}, 3.) ResNet-8 image classification on CIFAR-10 used in MLPerf\trademark{} tiny benchmark~\cite{mlperf}, and 4.) Novelty detection with OC-SVM~\cite{ocsvm}. Table~\ref{tab:workload_details} shows the peak performance characteristics of these benchmarks on the SoC, more specifically the RISC-V core and FlexML, with 8-bit precision, a single inference, and assuming all input data is available in the shared L2 memory. For TCN and ResNet-8, hardware-aware quantization was used and the energy and performance metrics were measured, while for the CAE and OC-SVM workloads, random inputs and weights were used. All the 4 workloads can be deployed with less than 230 $\mu$W of continuous real-time power at peak energy efficiency between 1-2 TOPS/W. This means that the SoC can provide high level of flexibility in workload mapping at sub-mW power to enable truly power efficient tinyML application on extreme edge devices. Fig.~\ref{fig_pb_bmarks} shows the power breakdown of the 4 real-time workloads. For OC-SVM (dense operation), the power consumption of memory dominates, due to the lack of re-usability of weights leading to more data fetches. On the other hand, power breakdown of CNN based workloads (TCN, ResNet8 and CAE) shows equal distribution between memory and logic as the dataflow exploits maximum re-usability.

\subsection{Power Management}

\begin{table}[!t]
\centering
\caption{Measurement results of different low power modes.}
\begin{tabular}{ccccc}
\toprule
\textbf{Power Mode} & \begin{tabular}[c]{@{}c@{}} \textbf{AON} \\ \textbf{Freq.} \\ \textbf{(kHz)} \end{tabular} & \begin{tabular}[c]{@{}c@{}}\textbf{Core}\\ \textbf{Freq.} \\ \textbf{(MHz)} \end{tabular} & \begin{tabular}[c]{@{}c@{}}\textbf{Power} \\ \textbf{($\mu$W)} \end{tabular} & \begin{tabular}[c]{@{}c@{}}\textbf{Wakeup} \\ \textbf{Latency} \\ \textbf{($\mu$s)}\end{tabular} \\ \midrule
Deep Sleep & 33 & - & 1.7 & 788  \\
\midrule
LP Data acq.$^*$ & 33 & 5 & 23.6 & 788  \\
\midrule
Data acq.$^*$ & 33 & 5 & 67 & 788  \\
\bottomrule
\multicolumn{4}{l}{$^{*}$ @$F_{s}$=44.1 kHz}
\end{tabular}%
\label{tab:pm_results}
\end{table}

\begin{figure}[!t]
\centering
\includegraphics[width=0.7\columnwidth]{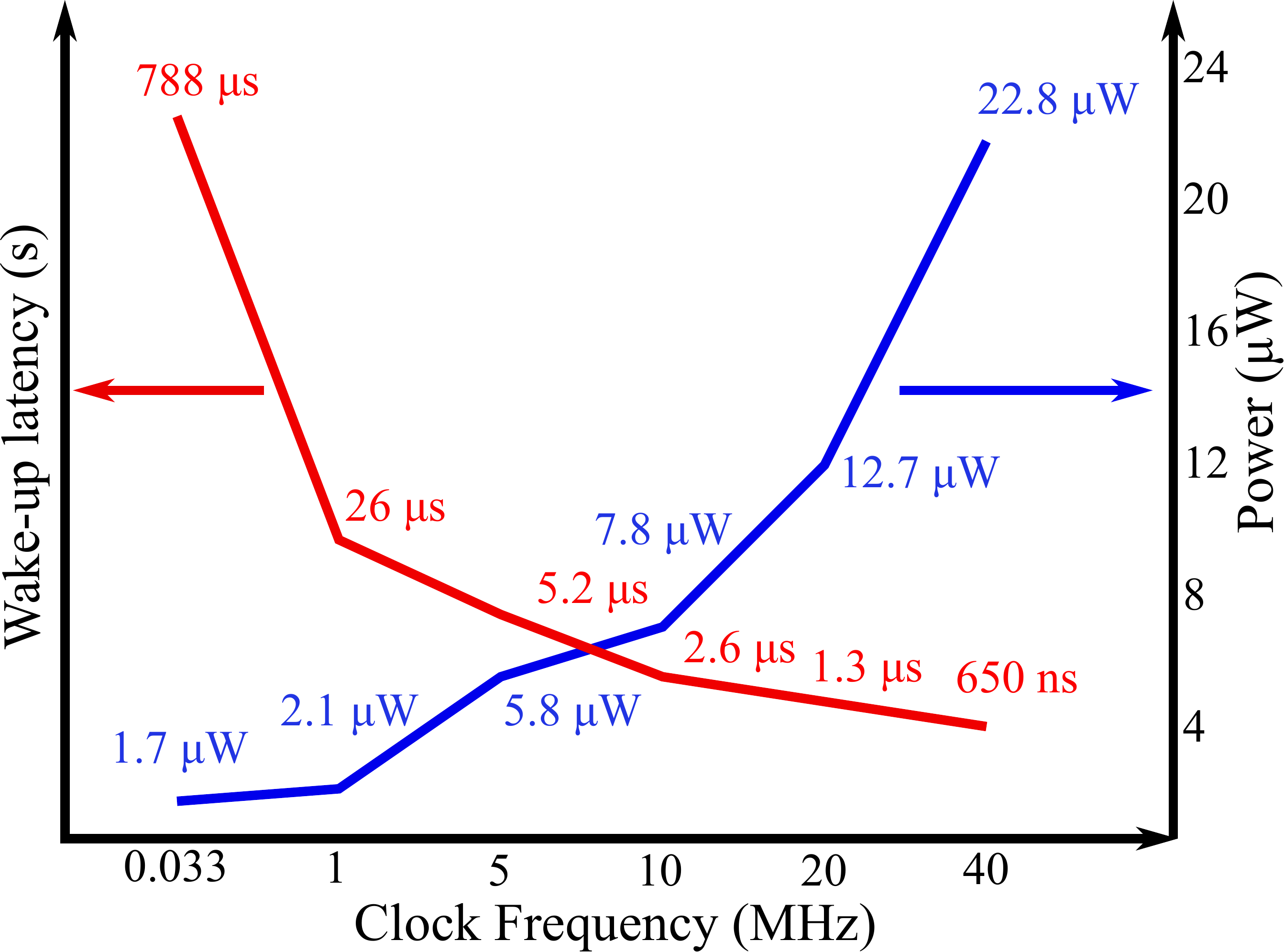}
\caption{Deep sleep power-latency-frequency tradeoff.}
\label{fig_14}
\end{figure}

Table~\ref{tab:pm_results} shows the measured real-time power of the different low power modes of the SoC detailed in Section~\ref{sec:pm}. In deep sleep mode the SoC operates with an AON clock frequency of 33 kHz. In this mode, only the AON domain consisting of the WuC and the logic controlling the IO pads stays powered ON. The resulting deep sleep power measured is 1.7 $\mu$W when operating at 0.7 V voltage supply. When compared to the peak power measured for the CNN layer, the deep sleep power is 12,000$\times$ lower. The measure latency of waking up the SoC from deep sleep mode to active mode is 788 $\mu$s. This wake-up latency can be traded off to deep sleep power by sweeping the AON clock frequency. Fig.~\ref{fig_14} plots the this trade-off for the measured power and wake-up latency when sweeping the AON clock frequency. Applications that need low latency can operate the AON clock at 40 MHz to attain a wake-up latency of 650 ns at a real-time power of 22.8 $\mu$W.


Table~\ref{tab:pm_results} also shows the measured power for the two tinyML optimized power modes of data acq. and LP data acq. These power modes are measured with an I2S protocol based windowed test vector collection with the AON clock frequency at 33 kHz and the core and peripheral clock frequency at 5 MHz. The SoC is programmed to collect I2S audio data through its uDMA at a sampling frequency of 44.1 kHz and a sampling window of 2 second. The sampling clock is generated by the SoC using the 5 MHz clock and lasts for the duration of sampling window. The data acq. or LP data acq. mode is then initiated and power is measured. 
The measured power for LP data acq. and data acq. is 23.6 $\mu$W and 67 $\mu$W resp. which is 850$\times$ and 300$\times$ reduced power consumption compared to the peak power, when the core and peripheral frequencies can be dynamically lowered to 5 MHz.  


\subsection{Instantaneous Power Trace}
\label{sec:inst_power}

In order to show the complete end-to-end application deployable on the SoC and to show the SoC's full ML functionality, duty cycling and features of power management, two applications are mapped onto the heterogeneous SoC with windowed data collection done in the LP data acq. mode: 
keyword spotting with a TCN model operating in continuous mode~\cite{tcn_work}; and a machine monitoring use case with a Mel Frequency Energy Coefficient (MFEC) based feature extraction with a CAE in duty cycled mode~\cite{cae}.

\begin{figure}[!t]
\centering
\includegraphics[width=\columnwidth]{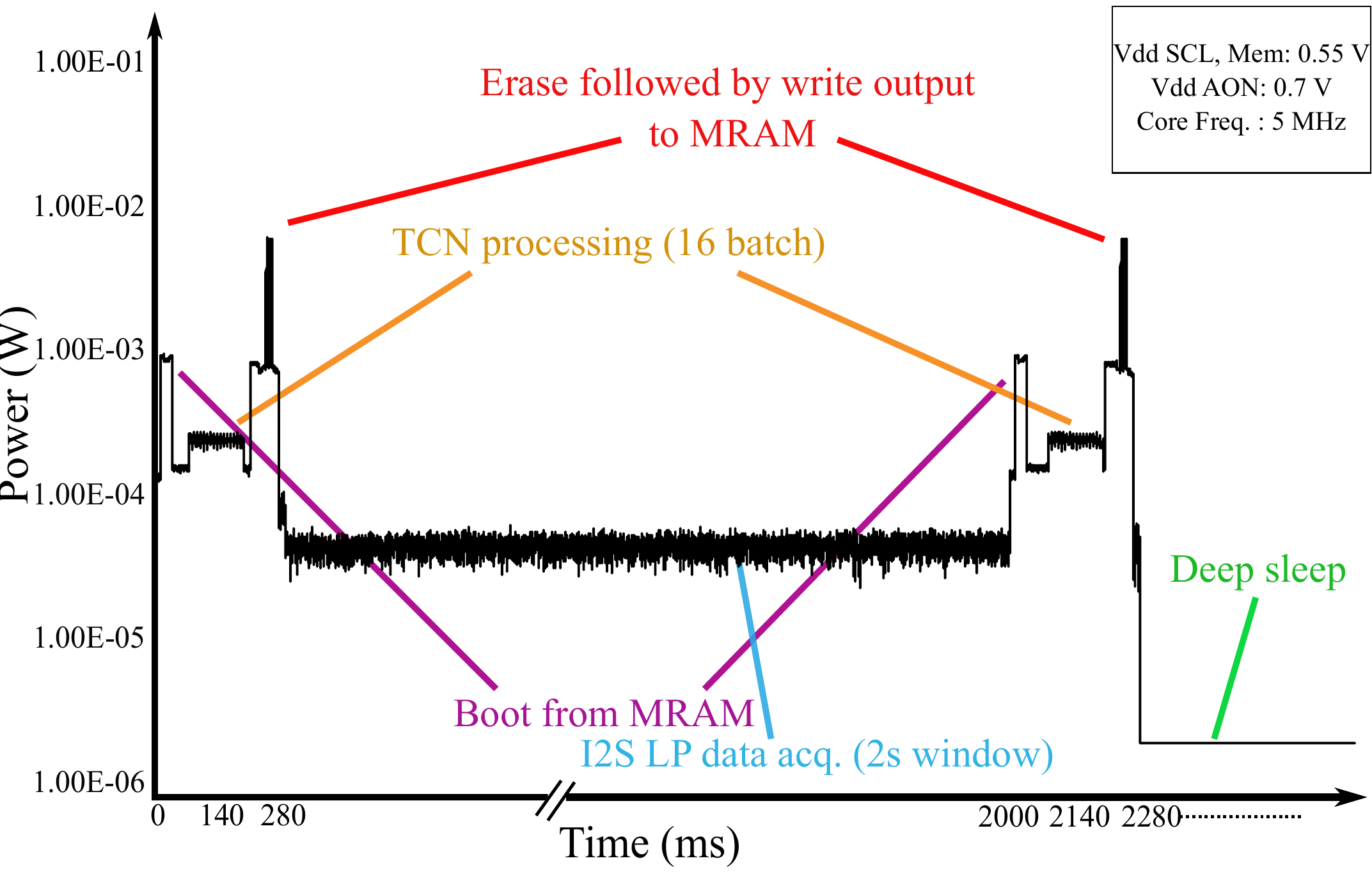}
\caption{Instantaneous power trace showing the KWS application scenario with one full period of smart sensing and TCN processing followed by idling.}
\label{fig_15}
\end{figure}

\begin{figure}[!t]
\centering
\includegraphics[width=\columnwidth]{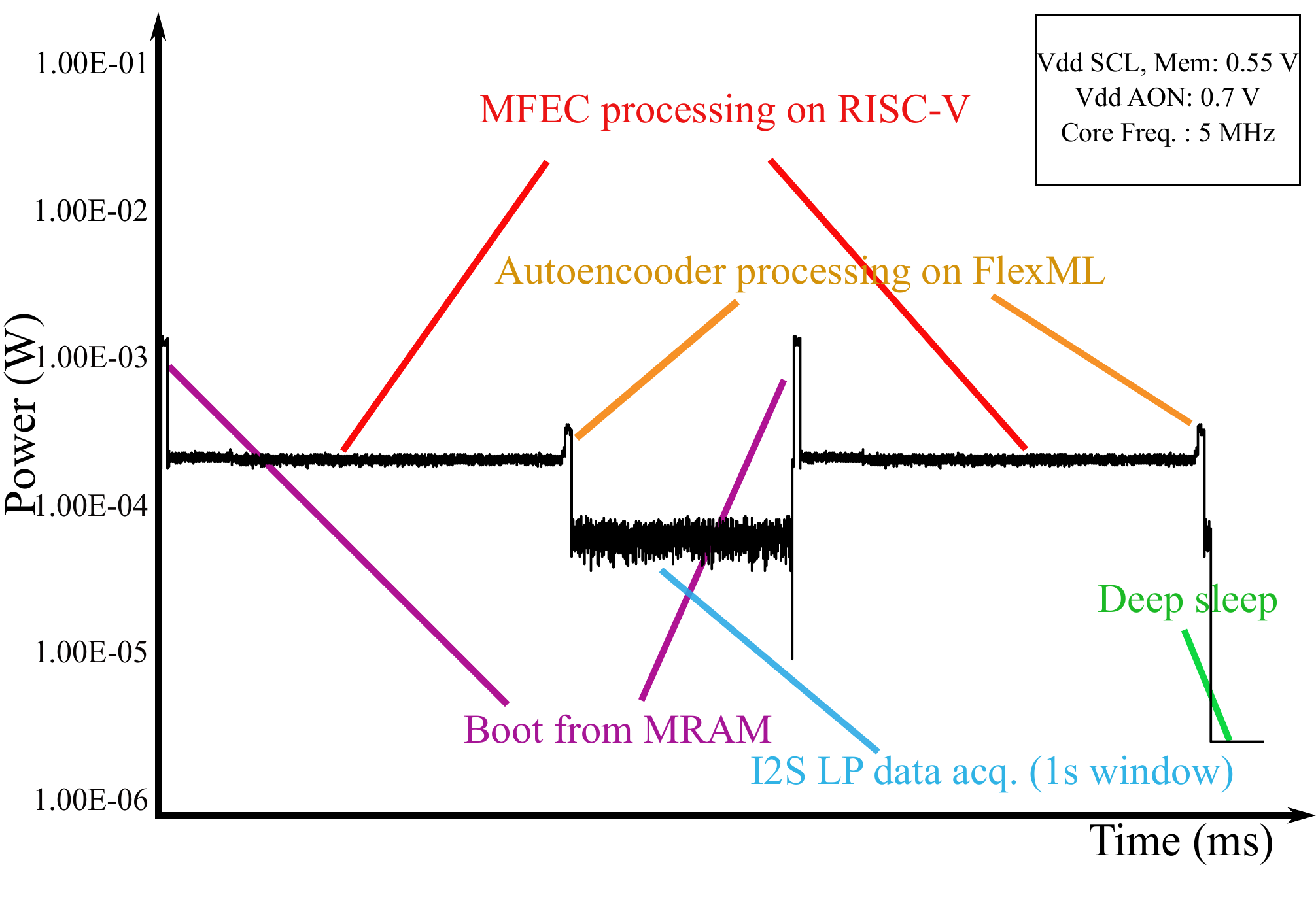}
\caption{Instantaneous power trace showing the machine monitoring application scenario with one period of smart sensing, FE, and CAE processing followed by idling.}
\label{fig_mfec_pt}
\end{figure}

\subsubsection{Keyword-spotting Application}
The first application scenario is the keyword-spotting with TCN model. In this application scenario, audio data from a microphone of window size 2 seconds (16 batches) at a sampling frequency of 44.1 kHz is collected using the I2S peripheral interface protocol, the collected data is simultaneously stored in the special L2 uDMA memory using the SoC's uDMA with the SoC being in the LP data acq. mode. After 2 seconds the SoC wake's up into active mode and the collected data is processed using the TCN model from Section~\ref{sec:ppa_workloads}. The output of the TCN processing is then stored into the MRAM for future processing or transmission while the SoC can either go into deep sleep mode or collect new windowed sampling data. Fig.~\ref{fig_15} shows the complete instantaneous power consumption trace of the KWS application scenario. When operating in this duty-cycled mode, the average power of the complete application is 173 $\mu$W. The power can be further reduced to 10-20 $\mu$W by using the deep sleep power mode of the SoC during periods of no sensing or computation. 


\begin{table*}[!t]
\centering
\caption{Performance comparison with state-of-the-art.}
\label{tab:sota}
\begin{tabular}{lccc|>{\columncolor{gray!20}}c|cc}
\toprule
 & \textbf{\cite{samurai}} & \textbf{\cite{vega}} & \textbf{\cite{gap}} & \textbf{TinyVers} & \textbf{\cite{shan}} & \textbf{\cite{unpu}}\\ \hline
 & & \multicolumn{1}{l}{\textbf{Extreme Edge SoCs}} & & & \multicolumn{2}{l}{\textbf{\hspace{0.6cm}edgeML Accelerators}} \\ \hline
\textbf{Technology}  & 28nm FDSOI & 22FDX & 55nm & 22FDX & 28nm & 65nm \\ \hline
\textbf{Die Area (mm$^2$)} & 4.5 & 12 & 10 & 6.25 & 0.55 & 16 \\ \hline
\textbf{Applications} & IoT GP, DNN,  & IoT GP, DNN,  & IoT GP, DNN, & IoT GP, DNN+, & Always-on KWS & DNN \\
 & NSA & NSA & NSA & Trad. ML, NSA & & \\ \hline
\textbf{Supported} & CNN, & CNN, & CNN,  & CNN, & DSCNN & CNN, \\
 \textbf{ML layers} & FC/RNN & FC/RNN & FC/RNN & FC/RNN, GAN, & & FC/RNN \\
 & & & &  AE, TCN, SVM & & \\ \hline
\textbf{Architecture} & 1$\times$RI5CY+ & 10$\times$RI5CY+ & 9$\times$RI5CY & 1$\times$RI5CY+ & DSCNN & DNN \\
 & ML accel. & ML accel. & & FlexML accel. & accel. & accel. \\ \hline
\textbf{SRAM } & 464 kB (40 kB) & 128 kB(L1) & 64 kB (L1) & 132 kB (L1) & 2 kB & 256 kB \\
\textbf{(State retentive)} & & (16-1600 kB (L2)) & (512 kB (L2)) & (64/512 kB (L2))  & &  \\ \hline
\textbf{eNVM} & - & 4 MB MRAM & - & 512 kB MRAM & - & - \\ \hline 
\textbf{Deep sleep power ($\mu$W)} & - & 1.7 & 3.6 & 1.7 & - & - \\ \hline
\textbf{SRAM ret.} & 6.4 & 2.8-123.7 & 30 & 23.6-67 & - & - \\
\textbf{sleep power ($\mu$W)} & & & & & & \\ \hline
\textbf{Int precision (bits)} & 8, 16, 32 & 8, 16, 32 & 8, 16, 32 & 2, 4, 8 & 8 & 1-16\\ \hline
\textbf{Supply voltage (V)} & 0.45-0.9 & 0.5-0.8 & 1-1.2 & 0.4-0.9 & 0.41 & 0.63-1.1 \\ \hline
\textbf{Max frequency (MHz)} & 350 & 450 & 250 & 150 & 0.04 & 200 \\ \hline
\textbf{Power range} & 6.4$\mu$W-96mW & 1.7$\mu$W-49.4mW & 3.6$\mu$W-75mW & 1.7$\mu$W-20mW & 0.51$\mu$W & 3.2-297mW\\ \hline
\textbf{Best ML perf.} & 36 GOPS & 32.2 GOPS & 12 GOPS & 17.6 GOPS  & 2.3 MOPS$^{+}$ & 691.2 GOPS  \\
 & @8b$^{*}$ & @8b$^{*}$ & @8b$^{*}$ & @8b$^{**}$ & @8b$^{**}$ & @8b $^{**}$\\ \hline
\textbf{Best ML eff.} & 1.3 TOPS/W@ & 1.3 TOPS/W@ & 200 GOPS/W@ & 2.47 TOPS/W@  & 4.5 TOPS/W@  & 5.57 TOPS/W, \\
 \textbf{@Perf} & 2.8 GOPS, 8b$^{*}$ & 15.6 GOPS, 8b$^{*}$ & 7 GOPS, 8b$^{*}$ & 0.58 GOPS, 8b$^{**}$ & 2.3 MOPS, 8b$^{**}$ & 8b$^{**}$\\
 &  &  &  & 11.9 TOPS/W@ &  & 11.6 TOPS/W, \\
 &  & &  & 2.4 GOPS, 2b$^{**}$ &  & 4b$^{**}$\\
\bottomrule
\multicolumn{4}{l}{$^{+}$ estimated at 90\% utilization of MACs, $^{*}$ Matmul, $^{**}$ CNN, 1 MAC = 2 Ops} \\

\end{tabular}%
\end{table*}

\subsubsection{Machine Monitoring Application} 
Machine monitoring used for predictive maintenance is the second application scenario selected. In this scenario, I2S peripheral interface protocol is used to collect audio data from a microphone with window size 1 second at a sampling frequency of 16 kHz. The collection of I2S audio data is operated in the LP data acq. mode of the SoC. Once the complete windowed data is collected, the SoC switches to the active mode in which the RISC-V core is used for the MFEC based feature extraction followed by running the CAE on the accelerator. 
Fig.~\ref{fig_mfec_pt} plots the instantaneous power trace of running the machine monitoring application. Unlike the previous application which works on raw audio data, the CAE model need pre-processing MFEC data. As the MFEC algorithm is not supported on the accelerator, it is executed on the RISC-V core with INT16 precision instead of INT32 or FP32 to reduce power consumption~\cite{mfec_approx}. The power trace plots show that running large feature extraction on RISC-V is not energy efficient 
taking large time to complete owing to single core operation. The average power for continuous operation remains below 164 $\mu$W, but for this use case, 9.5 $\mu$W is consumed with a duty cycling of 0.05. The MFEC execution on the RISC-V can be optimized using special DSP extensions available with the PULP libraries, which is left for future work. 

\section{Comparison with SoTA}
\label{sec:sota}
Table~\ref{tab:sota} shows the comparison of our SoC with SoTA on two fronts: on one hand, comparing with existing extreme edge SoCs (left), and on the other hand, with edge ML accelerators (right). Our SoC has similar or increased flexibility in application mapping compared to the extreme edge SoCs on the left, with much improved energy efficiency and power. TinyVers supports not only the IoT general processing (GP), DNNs and near-sensor analytics (NSA) like~\cite{samurai,vega,gap}, but also DNN+ such as TCN and AE and traditional ML like SVM, all at better energy efficiency because of efficient mapping. This is evident from the best energy efficiency of 2.47 TOPS/W for running a CNN layer on TinyVers. The energy efficiency is further enhanced to 11.9 TOPS/W when the CNN workload is quantized to INT2. Compared to the extreme edge SoCs, TinyVers provides support for precision scalability and, thus, can take advantage of improved performance using quantization. Furthermore, by utilizing support for block structured sparsity, TinyVers can reach a peak performance of 17 TOPS/W for an 8-bit CNN layer. This is much higher than the efficiencies reported by~\cite{samurai,vega,gap}. 

Compared to the edge ML accelerators on the right, TinyVers shows much more flexibility at comparable performance metrics in terms of energy efficiency and power consumption. The edgeML accelerators only support a single or few models extremely efficiently, but this approach has drawbacks in deployment for extreme edge devices. For example,~\cite{shan} can only perform KWS with depthwise separable CNN and its performance is much lower than TinyVers with comparable energy efficiency. UNPU~\cite{unpu}, can only support CNN and FC/RNN layers and also does not have a complete standalone SoC, which effects efficiency at the system level. Moreover, these edgeML accelerators cannot support any kind of duty cycling as they lack power management and retention memory support. TinyVers supports the multi-modal requirements of extreme edge devices at relatively similar energy efficiencies of the order of TOPS/W. Moreover, it adds the possibility of extreme low power idle states for duty-cycling use cases to enable $<10\mu$W operation, shown empirically in Section~\ref{sec:inst_power}. To summarize, TinyVers brings the best of both worlds of extreme edge processors and edgeML accelerators.


\section{Conclusion}
\label{sec:con}
TinyML applications at the extreme edge needs not only heterogeneous SoCs with flexible accelerators to support diverse workloads, but also adaptive power management for different duty-cycling operations. Moreover, to enable such adaptive power management, the need for embedded non-volatile memories arises. 
TinyVers extends a RISC-V core with a flexible ML accelerator supporting a diverse set of ML workload mapping in terms of diverse compute kernels, different precision and structured sparsity conditions. Furthermore, the inclusion of a WuC and an eMRAM enables the adaptive power management required in many duty-cycling use cases. Measurement result shows that the chip can achieve an energy efficiency range of 0.8-17 TOPS/W at 0.58 GOPS to 17.6 GOPS of throughput. The different low power modes enable the chip to achieve power range from 1.7$\mu$W-20 mW. The application of machine monitoring takes advantage of the deep sleep mode to consume only 9.5$\mu$W of power at a duty cycle of 0.05. Thus, TinyVers takes a step towards creating a new class of ultra-low power extreme edge SoCs.

\section*{Acknowledgments}
The authors would like to thank ETHZ for their support on PULP platform and GlobalFoundries for 22FDX tapeout support. The work has been supported under ISAAC project (FOD Economie Belgium Energietransitiefonds (oproep II)) in collaboration with Magics Technologies and received funding from the Flemish Government (AI Research Program).




\bibliographystyle{IEEEtran}
\bibliography{ref}

\clearpage

\newpage
\end{document}